\documentclass[fleqn,10pt]{wlscirep}

\usepackage[utf8]{inputenc}
\usepackage[T1]{fontenc}
\usepackage{amsmath,amssymb}
\usepackage{graphicx}
\usepackage{booktabs}
\usepackage{url}
\usepackage{xcolor}
\usepackage{hyperref}

\title{Bilinear gating of motor primitives: a principle linking dendritic computation to rapid goal-directed adaptation}

\author[1,2,*]{Cristiano Capone}
\author[1]{Luca Falorsi}
\author[1]{Andrea Ciardiello}
\author[3]{Luca Manneschi}

\affil[1]{Computational Neuroscience Unit, Istituto Superiore di Sanit\`a, 00161, Rome, Italy}
\affil[2]{Fondazione S.~Lucia IRCCS, Rome, Italy}
\affil[3]{School of Computer Science, University of Sheffield, Sheffield, S10 2TN, United Kingdom}
\affil[*]{cristiano0capone@gmail.com}

\begin{abstract}
Movement requires the motor cortex to specify both \emph{what} action to produce and \emph{which goal} it serves, yet how individual neurons separate these factors is not understood.
Here we show that in macaque motor cortex the \emph{burst fraction} of a neuron, the proportion of its spikes emitted in high-frequency bursts, encodes reach direction far more selectively than its overall firing rate.
This dissociation is highly consistent: it holds in every one of 12 recording sessions spanning three animals and two laboratories (all $p<10^{-12}$) and survives controls that remove any contribution of firing rate, showing that goal information is concentrated specifically in bursts.
We then show that this coding signature is the predicted consequence of dendritic coincidence detection in layer-5 pyramidal neurons: when a goal-related apical input coincides with a state-related basal drive the neuron bursts, so burst probability computes the product of goal and state, a bilinear gate $G(g)\,Y(s)$.
A minimal two-compartment spiking model reproduces the effect, and the same multiplicative gate, embedded in a reinforcement-learning agent, supports zero-shot generalisation to new goals and rapid online adaptation, providing a computational rationale for segregating goal information into bursts.
These results identify burst fraction as a goal-selective code in motor cortex, tie it to a concrete cellular mechanism, and show that the mechanism confers a learning advantage.
\end{abstract}

\begin{document}

\maketitle

\section{Introduction}

Flexible, goal-directed locomotion is a defining feature of vertebrate motor control. Animals shift between behavioral objectives (foraging, escape, social approach) within fractions of a second, yet the motor skills underlying each are acquired over days of experience. This dissociation implies that the nervous system maintains a repertoire of reusable motor primitives and recombines them rapidly according to the current goal, rather than storing an independent motor program for every possible objective~\cite{mussa1994combining,thoroughman2000learning}. Understanding the neural circuit mechanisms that implement this separation is a central question in motor neuroscience.

Layer~5 (L5) pyramidal tract neurons are anatomically and physiologically positioned to implement such a separation. Their basal dendrites integrate ascending sensorimotor inputs encoding body state, while their apical tufts receive descending projections from prefrontal and associative cortex encoding task context and goals~\cite{larkum2013thalamocortical,larkum1999dendritic}. Critically, the output mode of L5 neurons depends on whether both compartments are co-activated: basal drive alone produces tonic, regular spiking, while coincident apical and basal input triggers high-frequency burst discharge~\cite{larkum1999dendritic,hay2011division,urakubo2008nonlinear,shai2015}. This burst-coincidence mechanism has been proposed as a cellular AND-gate for contextual gating~\cite{capone2025adaptive,capone2023beyond}, but what specific computation it performs at the circuit level, and whether it is sufficient to implement flexible goal-directed control, remains an open question.

Inspired by this circuitry, we propose the bilinear form
\[
\mu(s,g) \;=\; \sum_{k=1}^{K} G_k(g)\,Y_k(s)
\]
as a general computational principle for goal-directed motor control, where $Y_k(s)$ are goal-independent motor primitives and $G_k(g)$ is a goal-dependent gating vector. We pursue three steps. First, we show that in motor cortex burst events encode the current goal far more selectively than tonic spikes, a prediction we confirm across 12 recording sessions from three animals, two institutions, and two task variants. Second, we show that this coding signature is the expected consequence of dendritic coincidence detection: a minimal population of two-compartment L5 neurons, in which burst probability computes the product $G_k Y_k$, reproduces the effect. Third, we show that the same multiplicative structure yields algorithmic advantages in modern deep reinforcement learning, through a bilinear actor-critic in which a shared gating vector $G$ simultaneously determines value estimates and the executed policy.

The proposed circuit has direct anatomical correlates. The thalamo-cortical loop~\cite{larkum2013thalamocortical,sherman2016thalamocortical} generates a high-dimensional dynamical embedding of sensory and internal signals that serves as sensorimotor working memory~\cite{maass2002real,jaeger2001echo}. Two parallel readout streams converge on L5 pyramidal neurons whose apical-basal coincidence detection drives motor output~\cite{rathelot2009motor}: a context-encoding pathway (prefrontal/L2/3) extracting slow goal variables~\cite{rigotti2013prefrontal,badre2009prefrontal} and a subpolicy pathway (premotor/SMA) computing state-dependent primitives~\cite{wise1985premotor,tanji2001sequential}. We demonstrate that this analogy is not merely metaphorical: a minimal population of two-compartment neurons physically realises the product $G_k Y_k$ at the single-cell level (Fig.~\ref{fig0}), and training such a network with reinforcement learning produces context-selective bursting and reliable multi-directional locomotion.

\section{Results}

A movement toward a goal requires the motor system to combine a state-dependent motor command with the currently active goal. We first ask how this combination is reflected in the spiking of individual motor cortex neurons, and in particular whether goal information is carried preferentially by burst events rather than by the overall firing rate, as predicted if goal context gates motor output through dendritic burst coincidence (a mechanism we develop below).

\begin{figure*}[t!]
\centering
\includegraphics[width=\linewidth]{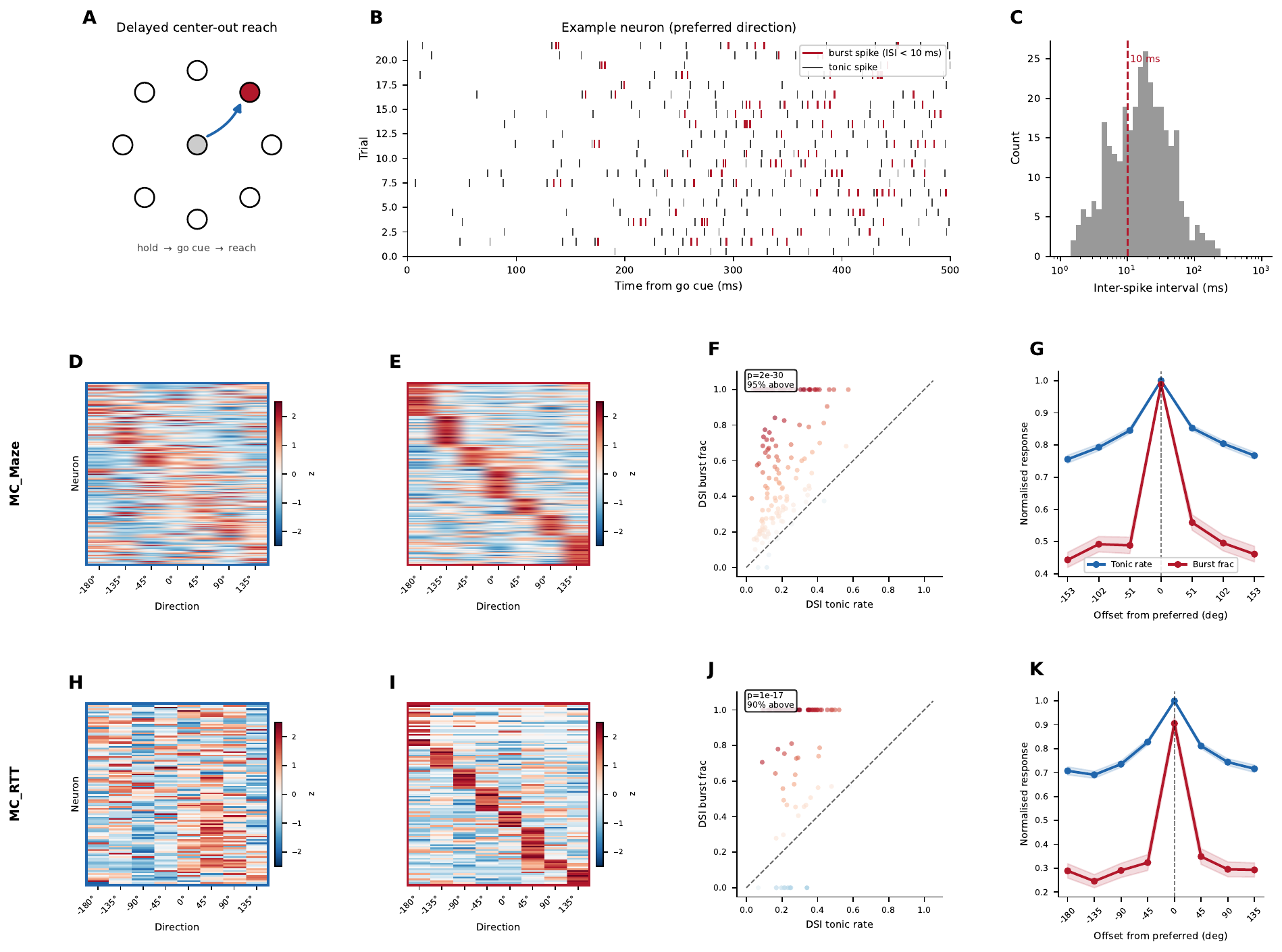}
\caption{\textbf{Burst fraction encodes reach direction more selectively than tonic spike rate.}
\textbf{(A)} Delayed center-out reaching task: after a hold and a go cue, the monkey reaches to one of several peripheral targets (red, active target).
\textbf{(B)} Spikes of an example neuron across trials to its preferred direction, aligned to the go cue (500\,ms window); each spike is classified as a burst spike if its interval to the preceding spike is below 10\,ms (red) or as tonic otherwise (grey).
\textbf{(C)} Inter-spike-interval distribution for the same neuron; the 10\,ms threshold (dashed) separates within-burst intervals from tonic firing.
\textit{(D-G)} MC\_Maze~\cite{churchland2012neural,pei2021nlb} (DANDI~000128; monkey Jenkins, Stanford; delayed centre-out reaching with maze barriers; $N=177$, 7 directions). \textit{(H-K)} MC\_RTT~\cite{pei2021nlb} (DANDI~000129; monkey Indy, UCSF; self-paced random target reaching; $N=106$, 8 directions).
\textbf{(D,H)} Population activity matrices for tonic rate (neurons sorted by burst-fraction preferred direction, z-scored).
\textbf{(E,I)} Same for burst fraction: a pronounced diagonal staircase is visible in both datasets, absent from the tonic rate matrices.
\textbf{(F,J)} Per-neuron DSI for tonic rate vs.\ burst fraction. Burst fraction DSI is significantly higher in both datasets (MC\_Maze: $p<10^{-29}$, 95\% above diagonal; MC\_RTT: $p<10^{-16}$, 90\% above diagonal; Wilcoxon signed-rank test).
\textbf{(G,K)} Mean population tuning curves ($\pm$\,SEM) aligned to each neuron's preferred direction, confirming genuine peaked tuning in both signals.}
\label{fig0b}
\end{figure*}

\paragraph{Neural evidence: burst fraction encodes reach direction in motor cortex.}
The burst coincidence model predicts that directional information in motor cortex should be disproportionately concentrated in burst events rather than total spike counts.
An example neuron illustrates the distinction: aligned to its preferred reach direction, its spikes separate into burst spikes (inter-spike interval below 10\,ms to the preceding spike) and tonic spikes (Fig.~\ref{fig0b}A-C). For each neuron we computed the mean tonic spike rate and burst fraction in the 500\,ms post-movement-onset window.
We tested the prediction on two independent M1/PMd datasets from different animals, labs, and task variants (Fig.~\ref{fig0b}).

The first dataset is MC\_Maze~\cite{churchland2012neural,pei2021nlb} (DANDI~000128; monkey Jenkins, Stanford; delayed centre-out reaching with maze barriers; $N=177$ neurons, 7 directions; Fig.~\ref{fig0b}D-G).
Population activity matrices sorted by burst-fraction preferred direction reveal a pronounced diagonal staircase in burst fraction (Fig.~\ref{fig0b}E) that is absent from the tonic rate matrices sorted on the same neuron ordering (Fig.~\ref{fig0b}D).
Burst fraction DSI is significantly higher than tonic rate DSI across the population (Fig.~\ref{fig0b}F; Wilcoxon signed-rank test, $p < 10^{-29}$, 95\% of neurons above diagonal).

The second dataset is MC\_RTT~\cite{pei2021nlb} (DANDI~000129; monkey Indy, UCSF; self-paced random target reaching, no delay; $N=106$ neurons, 8 directions; Fig.~\ref{fig0b}H-K).
The same pattern replicates in this entirely independent recording: burst fraction again shows a clear directional staircase while tonic rate does not, and burst fraction DSI is significantly higher ($p < 10^{-16}$, 90\% of neurons above diagonal).

Population tuning curves aligned to each neuron's preferred direction (Fig.~\ref{fig0b}G,K) confirm genuine peaked tuning in both signals in both datasets, ruling out that the higher burst-fraction DSI reflects noise.
To assess replicability, we extended the analysis to 11 additional sessions from the same two animals recorded across multiple days (DANDI~000070~\cite{churchland2012neural}), as summarised next.

\section{Replication across 12 sessions, three animals, and two institutions}
\label{sec:replication}

The effect replicated in every qualifying session: across all 12 sessions (three animals, two institutions, two task variants), 69 to 95\% of neurons showed higher burst-fraction than tonic-rate direction selectivity, with all $p<10^{-12}$ (Wilcoxon signed-rank test; Table~\ref{tab:sessions}).
Two controls confirm the effect is not a trivial consequence of firing rate. A bootstrap over neurons (200 resamples) placed the percentage above diagonal well above chance in every session, and the difference persisted when burst-fraction selectivity was recomputed on the rate-residual burst fraction, that is the burst fraction after regressing out tonic rate across directions (MC\_Maze 99\% above diagonal, $p=4.5\times10^{-31}$; MC\_RTT 91\%, $p=9.4\times10^{-19}$).
Sessions from monkey Nitschke, which had unbalanced direction sampling, were restricted to direction bins with at least 30 trials (4 to 7 of 8 bins retained); full datasets, spike extraction, binning, and statistical procedures are given in Methods.

\begin{table}[h!]
\centering
\caption{\textbf{Replication across sessions, animals, and institutions.}
DSI = direction selectivity index; \% above = fraction of neurons with burst-fraction DSI $>$ tonic-rate DSI; $p$ = Wilcoxon signed-rank test (one-sided, burst frac $>$ tonic rate).
Nitschke sessions used direction bins with $\geq 30$ trials; the number of retained bins is shown in parentheses.}
\label{tab:sessions}
\small
\begin{tabular}{llllrrl}
\toprule
Animal & Institution & DANDI & Session & $N$ & \% above & $p$ \\
\midrule
Jenkins & Stanford & 000128 & 20090925 & 177 & 95\% & $2.3\times10^{-30}$ \\
Jenkins & Stanford & 000070 & 20090912 & 191 & 94\% & $1.1\times10^{-31}$ \\
Jenkins & Stanford & 000070 & 20090916 & 191 & 94\% & $2.9\times10^{-32}$ \\
Jenkins & Stanford & 000070 & 20090918 & 192 & 95\% & $5.8\times10^{-32}$ \\
Jenkins & Stanford & 000070 & 20090923 & 191 & 93\% & $1.1\times10^{-31}$ \\
\midrule
Nitschke & Stanford & 000070 & 20090812 (6/8 dirs) & 191 & 91\% & $1.7\times10^{-28}$ \\
Nitschke & Stanford & 000070 & 20090819 (6/8 dirs) & 192 & 85\% & $1.2\times10^{-25}$ \\
Nitschke & Stanford & 000070 & 20090910 (7/8 dirs) & 191 & 76\% & $1.1\times10^{-18}$ \\
Nitschke & Stanford & 000070 & 20090920 (4/8 dirs) & 191 & 74\% & $8.8\times10^{-16}$ \\
Nitschke & Stanford & 000070 & 20090922 (4/8 dirs) & 192 & 85\% & $1.2\times10^{-23}$ \\
Nitschke & Stanford & 000070 & 20100923 (7/8 dirs) & 192 & 69\% & $4.4\times10^{-13}$ \\
\midrule
Indy    & UCSF     & 000129 & 20170202             & 106 & 90\% & $1.3\times10^{-17}$ \\
\bottomrule
\end{tabular}
\end{table}

\begin{figure*}[t!]
\centering
\includegraphics[width=\linewidth]{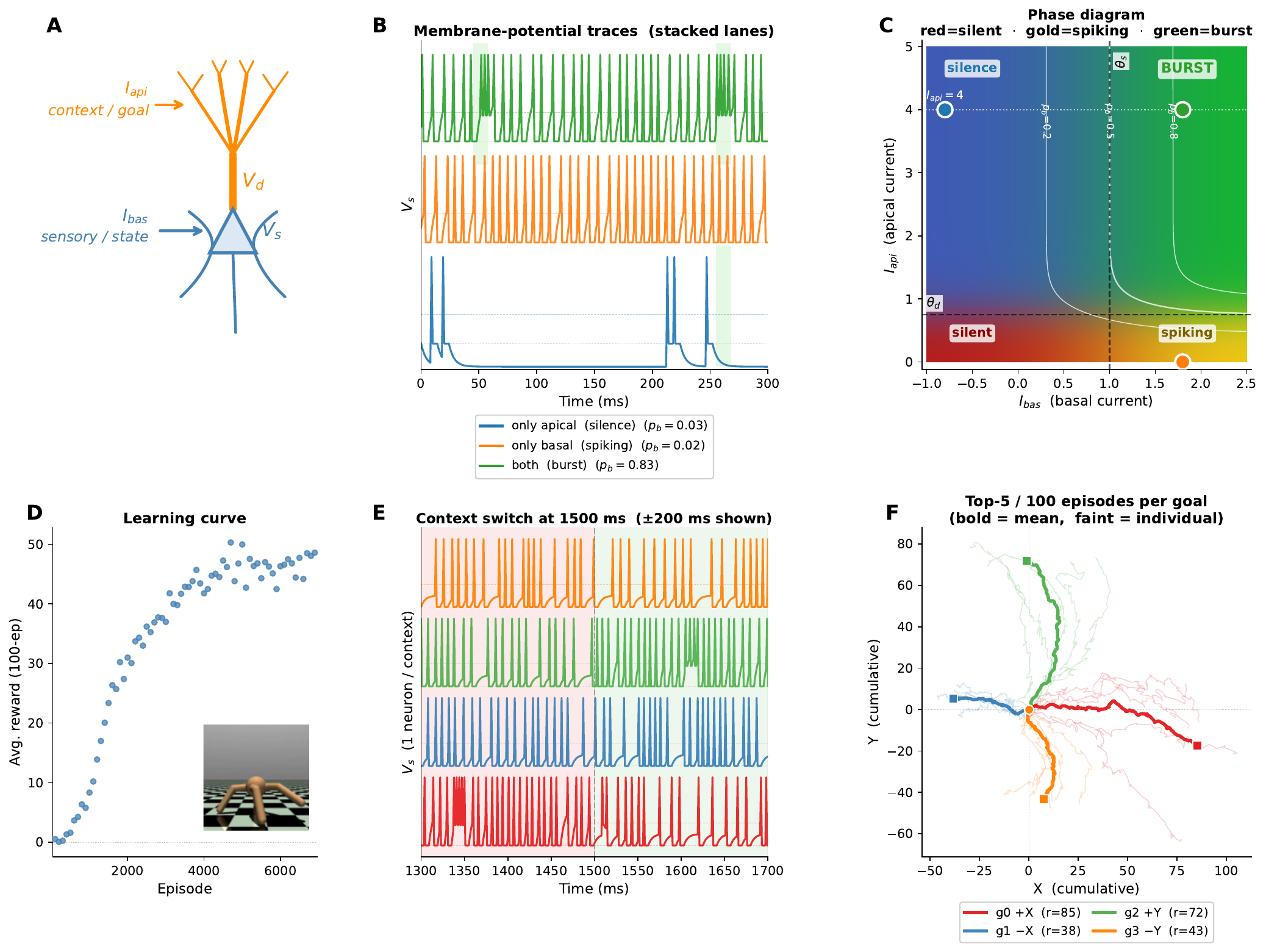}
\caption{\textbf{Burst-gated neural actor with voltage-based burst probability for goal-conditioned locomotion.}
\textbf{(A)} Two-compartment pyramidal neuron model. Feedforward basal current $I_\text{bas}$ drives the somatic voltage $V_s$; top-down apical current $I_\text{api} = g \cdot I_\text{API\_SCALE}$ ($I_\text{API\_SCALE}=4.0$) drives the dendritic voltage $V_d$. Burst requires co-activation of both compartments.
\textbf{(B)} Somatic membrane-potential traces ($V_s$, 300\,ms) under three input conditions (stacked lanes). Blue: apical only ($I_\text{api}=4.0$, $I_\text{bas}=-0.8$): gate open but soma sub-threshold, near-silence ($p_\text{burst}=0.03$). Orange: basal only ($I_\text{bas}=1.8$, $I_\text{api}=0$): regular spiking, burst gate closed ($p_\text{burst}=0.02$). Green: both ($I_\text{bas}=1.8$, $I_\text{api}=4.0$): dense recurrent bursting ($p_\text{burst}=0.83$, green shading marks burst windows).
\textbf{(C)} Two-dimensional phase diagram of the burst mechanism using a bilinear RGB colormap encoding $p_s(I_\text{bas})$ and $p_d(I_\text{api})$ independently. Four regions are visually separated: red (silent, $p_s\approx0$, $p_d\approx0$), blue (apical-only silence, $p_s\approx0$, $p_d\approx1$), gold (spiking without burst, $p_s\approx1$, $p_d\approx0$), and green (burst, $p_s\approx1$, $p_d\approx1$). Contours show $p_\text{burst} \in \{0.2,\,0.5,\,0.8\}$. Dashed lines mark $\theta_s=1.0$ and $\theta_d=0.75$; dotted white line marks the active apical drive $I_\text{api}=4.0$. Coloured dots correspond to the three conditions in~(B).
\textbf{(D)} Learning curve for the four-goal MuJoCo Ant-v4 task (100-episode moving average). The agent converges to average rewards above 30 within ${\sim}$3500 episodes. Inset: screenshot of the simulated ant.
\textbf{(E)} Post-training $V_s$ traces (stacked lanes, one neuron per goal context) showing 200\,ms before and 200\,ms after the context switch at $t=1500$\,ms ($100~\text{steps} \times 15~\text{ms}$). Goal~0 (red) bursts densely in the preceding window and stops abruptly at the switch; goal~2 (green) transitions from sub-threshold silence into sustained bursting. Goals~1 and~3 remain sub-threshold throughout, confirming context selectivity.
\textbf{(F)} Top-5 out of 100 evaluation episodes per goal direction (faint = individual, bold = mean; circles = start, squares = end). Mean rewards: 85 ($+X$), 38 ($-X$), 72 ($+Y$), 43 ($-Y$).}
\label{fig0}
\end{figure*}

Having established that burst fraction carries the directional signal in vivo, we asked whether a biophysically grounded mechanism can produce it: a population of two-compartment neurons in which a goal-selective apical signal and a state-dependent basal drive interact through burst coincidence, so that burst probability computes their product (Fig.~\ref{fig0}).

\paragraph{Voltage-based burst probability enables context-selective motor gating.}
We implemented the burst actor using a voltage-based burst probability formulation (V-prob), in which the motor output is the time-averaged product of two sigmoid-transformed membrane voltages across each 15-ms action window:
\begin{equation}
p_\text{burst} = \bigl\langle p_s(V_s(t))\cdot p_d(V_d(t))\bigr\rangle_t, \qquad
p_s = \sigma\!\left(\tfrac{V_s-\theta_s}{\beta_s}\right), \quad
p_d = \sigma\!\left(\tfrac{V_d-\theta_d}{\beta_d}\right).
\label{eq:pburst_results}
\end{equation}
The motor command is then $\mu = P_\text{scale}\cdot p_\text{burst} - 1$ (full parameter specification in Methods). The core property of this formulation is that $p_\text{burst} \approx 0$ unless both compartments are simultaneously above threshold, directly instantiating the bilinear gate $G_k Y_k$ at the single-neuron level: $p_d$ encodes the goal-context gate $G_k$ and $p_s$ encodes the state-dependent drive $Y_k$. The apical drive strength was set so that $p_d \approx 1$ whenever the corresponding goal is active, ensuring burst probability faithfully tracks the somatic response.

\paragraph{Phase structure of the burst gate.}
The two-dimensional phase diagram (Fig.~\ref{fig0}C) reveals four functionally distinct regions by independently encoding the somatic activation $p_s(I_\text{bas})$ and the dendritic gate $p_d(I_\text{api})$ through a bilinear RGB colormap. The silent region (red, low $I_\text{bas}$, low $I_\text{api}$) and apical-only silence region (blue, high $I_\text{api}$, low $I_\text{bas}$) confirm that strong apical drive alone is insufficient to activate the neuron when the soma is sub-threshold. The spiking region (gold, high $I_\text{bas}$, low $I_\text{api}$) shows tonic somatic firing without burst mode. Only in the burst region (green, upper-right), where the learned basal drive brings $V_s$ above $\theta_s$ and the goal context elevates $V_d$ above $\theta_d$ simultaneously, does the neuron enter the rapid-bursting regime. With $I_\text{API\_SCALE} = 4.0$, the active operating point (white dotted line at $I_\text{api} = 4.0$) lies far above $\theta_d = 0.75$, ensuring the gate is fully open with essentially zero probability of accidental closure during an active context (Fig.~\ref{fig0}B-C).

\paragraph{Context-selective bursting and abrupt context switching.}
As shown in Fig.~\ref{fig0}B, only when both $I_\text{bas}$ and $I_\text{api}$ are simultaneously elevated does the neuron produce burst events (green shading, $p_\text{burst}=0.83$); all other conditions result in silence or regular tonic spiking. The post-training context-switching experiment (Fig.~\ref{fig0}E), zoomed to $\pm$200\,ms around the switch at $t=1500$\,ms, demonstrates abrupt and complete selectivity: goal-0 neurons cease bursting within one burst-refractory cycle (${\sim}$45\,ms) of the context transition, and goal-2 neurons enter sustained bursting within the same interval. Goals~1 and~3, which receive no apical drive throughout the 400-ms window, remain sub-threshold, confirming that burst activity is exclusively driven by coincident basal-apical activation.

\paragraph{Learning and navigation.}
A key biological feature of this formulation is that the learning rule is both local and online. Because $p_\text{burst}$ depends only on the neuron's own membrane voltages, the policy gradient with respect to the readout $y$ reduces to a three-factor Hebbian rule:
\begin{equation}
\Delta w_j \;\propto\; r(t)\cdot x_j(t)\cdot \frac{\partial\mu}{\partial y}(t), \qquad
\frac{\partial\mu}{\partial y} = P_\text{scale}\cdot p_\text{burst}(1-p_\text{burst})\cdot I_\text{scale}\cdot\operatorname{sech}^2(y)\,/\,\beta_s,
\label{eq:local_rule}
\end{equation}
where $r(t)$ is the instantaneous reward, $x_j(t)$ is the presynaptic input, and $\partial\mu/\partial y$ depends only on the neuron's own somatic state via $p_\text{burst}$. Each synapse is therefore updated using only locally available quantities plus a global reward broadcast, with no backpropagation across the network (full derivation in Methods).

The V-prob agent learns to navigate in all four cardinal directions (Fig.~\ref{fig0}D-F). The learning curve shows a consistent increase in episode reward, reaching a plateau around 30-35 by episode 3500. Evaluation over 100 episodes per goal yields mean top-5 rewards of 85, 38, 72, and 43 for goals $+X$, $-X$, $+Y$, and $-Y$ respectively. While lower than the original equilibrium-based formulation, the V-prob agent still produces directionally selective trajectories in all four goals; the stronger apical gate ($I_\text{API\_SCALE} = 4.0$, $p_d \approx 1$ when active) guarantees sharper context disambiguation at the neural level, suggesting that the reduced reward reflects a training dynamics difference rather than a failure of the gating mechanism.

\begin{figure*}[t!]
\centering
\includegraphics[width=0.9\linewidth]{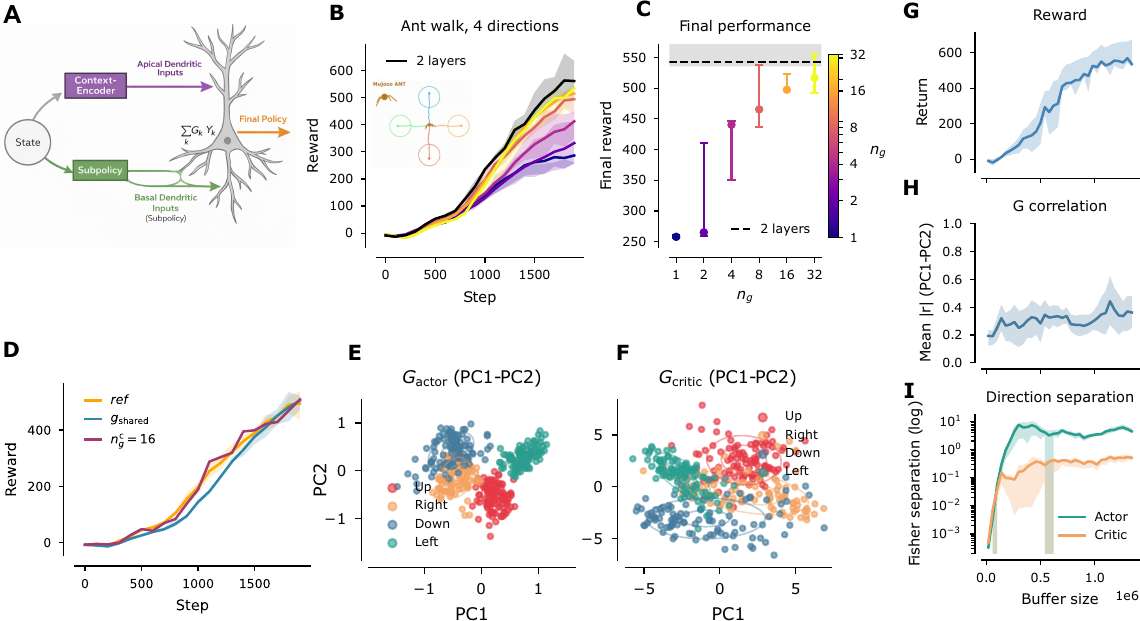}
\caption{\textbf{MLP-based bilinear actor-critic architecture.}
\textbf{A.} Scheme of our architecture: the actor and critic are decomposed into $K$ parallel basis modules, policy primitives $Y_k(s)$ and value components $\psi_k(s,a)$.
\textbf{B.} Biological interpretation: sensory and contextual inputs affect the dynamics of pyramidal neurons in a multiplicative way.
\textbf{C.} Scheme of the navigation task: a robot with 8 DOF is asked to move in a specific direction.
\textbf{D-E.} Comparison between the learning curves of a traditional architecture (2-layer MLP) and a single-layer MLP with bilinear decomposition.
\textbf{F.} Comparison of learning curves between the cases in which the latent space $G_k$ is independent or shared between actor and critic.
\textbf{G.} Direction encoding for actor and critic.
\textbf{H-I.} Reward, correlation between actor and critic $G$, and direction encoding in the $G$ space, as functions of training steps.}
\label{fig1}
\end{figure*}

Part~I established that the bilinear product $G_k Y_k$ can be physically implemented by spiking neurons (Fig.~\ref{fig0}) and that burst fraction carries the predicted directional signal in motor cortex recordings (Fig.~\ref{fig0b}). Part~II asks a separate question: does the multiplicative structure itself, independently of any spiking implementation, yield algorithmic advantages when embedded in a modern deep RL framework? We instantiate the bilinear decomposition as a standard differentiable actor-critic trained with Soft Actor-Critic (SAC), which provides the sample efficiency and stability required for continuous high-dimensional action spaces in simulated robotic environments. The central RL advantage under investigation is the shared gating vector $G$: because actor and critic are both linear in the same $G$, updating $G$ to fit a new reward structure simultaneously reshapes the executed policy, removing the separate policy gradient step that standard methods require. This is a claim about the \emph{structure} of the adaptation problem, not about asymptotic reward; accordingly, we evaluate it by ablation (sharing $G$ vs.\ learning actor and critic gates independently, holding everything else fixed) and by adaptation cost (the number of policy-improvement steps eliminated), rather than by a head-to-head reward benchmark against methods that solve a different adaptation problem. Panel~A of Fig.~\ref{fig1} illustrates the shared-coefficient architecture; panel~B provides the biological interpretation; panel~C defines the task setting.

The main quantitative result is that bilinear decomposition improves learning efficiency even with a shallower network. As shown in Fig.~\ref{fig1}D-E, a single-layer bilinear model reaches higher reward faster than a standard two-layer MLP baseline, indicating that multiplicative structure can compensate for depth by providing a more task-aligned representation.

We then test the central design choice of this work: sharing the gating coefficients between actor and critic. Fig.~\ref{fig1}F shows that using a common latent vector $G$ yields performance comparable to (or better than) the variant with separate actor/critic gates, while reducing parameterization and optimization complexity.

Finally, Fig.~\ref{fig1}G-I characterizes the learned latent space. The $G$ coordinates encode movement direction in a structured way, and the actor-critic gating vectors become increasingly correlated during training as reward improves. These trends support our claim that shared $G$ forms a coherent control interface that is both behaviorally meaningful and suitable for rapid adaptation.

We empirically investigated the effect of sharing the $G$ components between the actor and the critic in our proposed bilinear decomposition framework, where
\[
Q(s,a) = \sum_k G_k(s) \, V_k(a,s), \quad \mu(s) = \sum_k G^{(A)}_k(s) \, Y_k(s).
\]
In the general formulation, the actor and the critic may have different $G$ functions ($G^{(A)} \neq G$), allowing them to be optimized independently. However, our experiments show that setting $G^{(A)} = G$ leads to essentially identical performance compared to learning $G^{(A)}$ separately.

This finding has two important consequences:
\begin{enumerate}
    \item It is not necessary to optimize $G^{(A)}$ at all; one can directly reuse the critic's $G$ for the actor, thus reducing the number of parameters and simplifying the optimization process.
    \item When facing a new task, adaptation can be achieved by re-estimating only $G$ from the critic, without re-learning $G^{(A)}$ and subsequently re-optimizing the actor. This substantially reduces the cost of transfer to new tasks.
\end{enumerate}

To our knowledge, this is the first actor-critic in which the actor reuses the critic's $G$ exactly, making policy improvement a free byproduct of value estimation.
We emphasise that the contribution here is \emph{structural} rather than a performance improvement: the relevant test is not whether shared-$G$ outperforms a stronger baseline on raw reward, but whether coupling actor and critic through a single $G$ removes the separate policy-improvement step \emph{without} sacrificing performance.
The controlled comparison for this claim is therefore the ablation between shared and independent gating (Fig.~\ref{fig1}F; Fig.~\ref{fig5}, GShares vs.\ GXY), which holds the architecture fixed and varies only whether $G$ is shared, and which shows that sharing incurs no performance cost while eliminating the actor-side optimisation entirely.

\begin{figure*}[t!]
\centering
\includegraphics[width=0.9\linewidth]{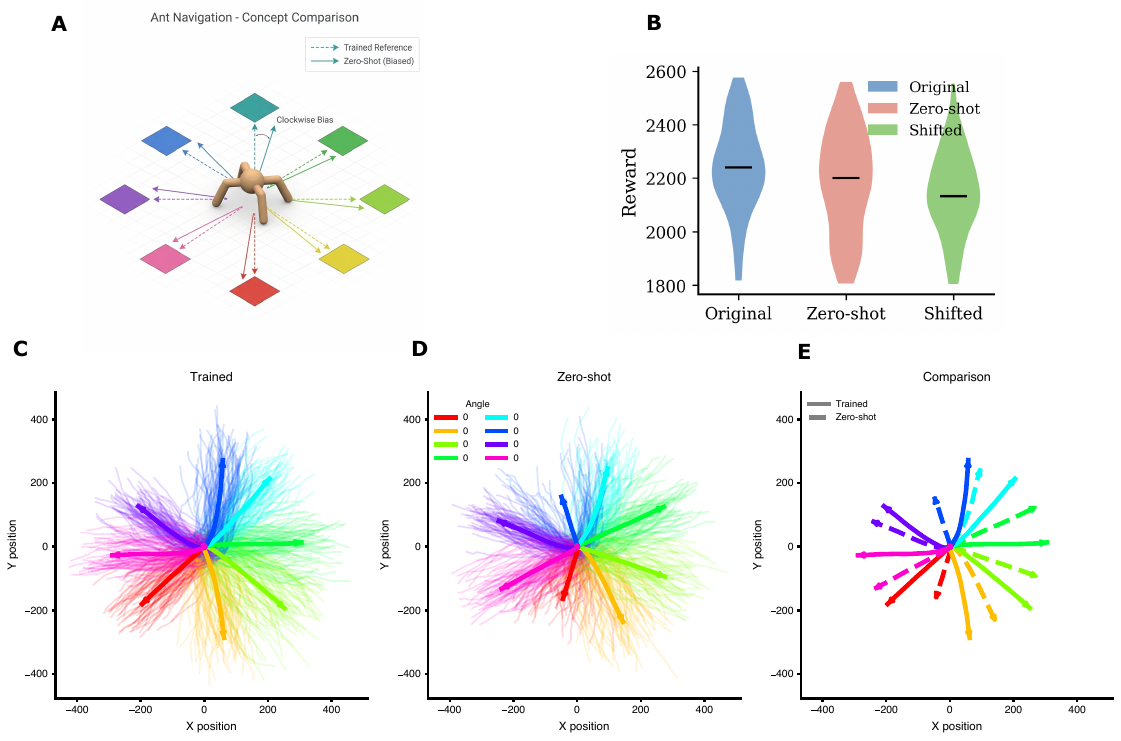}
\caption{\textbf{Zero-shot learning.} \textbf{A.} Task scheme: the MuJoCo Ant agent is pre-trained on target directions and tested on new ones. Pretrained bilinear agent is evaluated on unseen goal directions (or task descriptors) without any parameter update, by conditioning on $g$. \textbf{B.} Performance compared against baselines when switching to novel directions. \textbf{C-D.} Behavior trajectories for training and test directions, respectively, illustrating successful generalization to intermediate angles not explicitly trained. \textbf{E.} Direct comparison between train and test directions (averaged over trials).}
\label{fig2}
\end{figure*}

We next evaluate zero-shot generalization to unseen target directions (Fig.~\ref{fig2}). Panel A summarizes the protocol: the MuJoCo Ant is pretrained only on directional objectives and then tested on new directions without any parameter update, conditioning only on $g$. Panel B shows that the pretrained bilinear agent remains competitive with baseline methods after direction switches. Panels C-D report representative trajectories for training and test directions, showing smooth interpolation toward intermediate headings that were not explicitly seen during training. Importantly, this zero-shot transfer is not limited to heading selection: modulation in $G$ also produces coherent changes in movement speed, even though speed was never an explicit training target. Finally, panel E compares train and test directional performance averaged over trials, confirming only limited degradation in the zero-shot regime and supporting the view that $G$ acts as a structured control variable rather than a standard contextual input.

\begin{figure*}[t!]
\centering
\includegraphics[width=\linewidth]{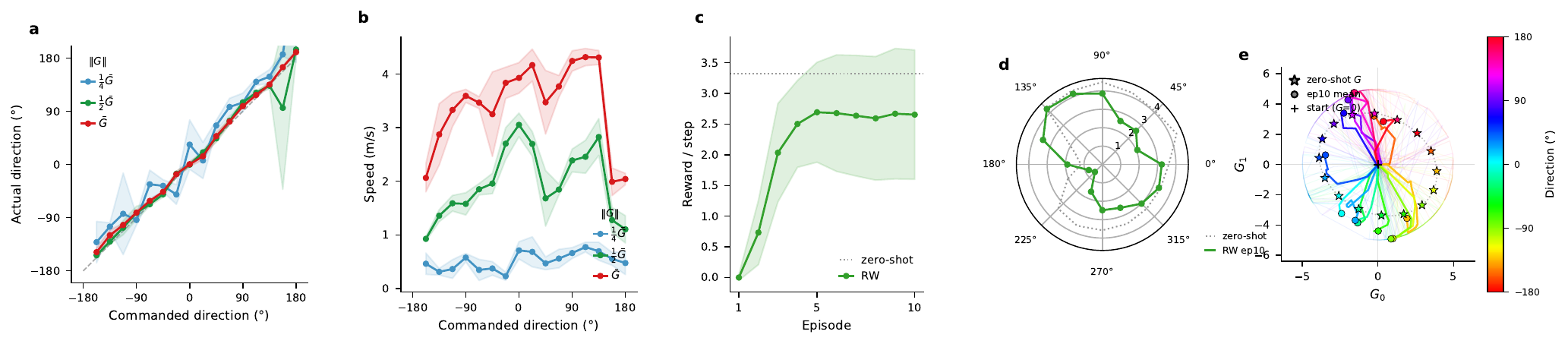}
\caption{\textbf{Zero-shot directional control and reward-weighted cold-start adaptation.}
\textbf{(A)} Actual movement direction as a function of commanded direction for three $G$ magnitudes ($\tfrac{1}{2}\bar{G}$, $\bar{G}$, $2\bar{G}$, where $\bar{G}$ is the mean zero-shot norm). Across all amplitudes, the ant tracks the commanded direction closely (dashed line: identity), demonstrating that the direction of $G$ controls movement direction independently of its magnitude.
\textbf{(B)} Locomotion speed as a function of commanded direction for the same three amplitudes. Speed is modulated by $\|G\|$ while remaining approximately uniform across directions, confirming that $G$ magnitude and direction act as independent control axes for speed and heading.
\textbf{(C)} Cold-start adaptation with reward-weighted $G$ updates (RW). Starting from $G=0$, the agent recovers behaviorally useful $G$ vectors within a few episodes. Mean reward per step ($\pm$ s.d.\ across 16 directions) is shown; the dotted line indicates the zero-shot performance level.
\textbf{(D)} Polar plot of reward per step at episode 10 as a function of movement direction, for the zero-shot policy (dotted) and after RW adaptation (solid). Adaptation improves performance across all directions.
\textbf{(E)} Trajectories of the $G$ vector in the $G_0$-$G_1$ plane during RW adaptation. Stars mark the zero-shot $G$ positions computed by the pretrained $g$-layer; circles mark the mean adapted $G$ after 10 episodes. Arrows indicate the direction of adaptation; the dotted curve shows the zero-shot locus across all commanded directions.}
\label{fig3}
\end{figure*}

\paragraph{Zero-shot directional control.}
The pretrained model generalises immediately to all 16 evaluated directions, including 8 not seen during training, achieving a mean reward of $3.4 \pm 0.4$ r/step across directions and episodes without any test-time adaptation (Fig.~\ref{fig3}C, dotted). The $g$-layer encodes the task context $(\cos\theta, \sin\theta)$ through a linear map whose image traces a smooth ellipse in $G$-space (Fig.~\ref{fig3}E), providing continuous interpolation between discrete training directions.

\paragraph{Decoupled control of heading and speed.}
We verify that the direction and magnitude of $\mathbf{G}$ constitute independent control axes by fixing the direction of $\mathbf{G}$ to that given by the zero-shot map and scaling its norm to $\tfrac{1}{2}\bar{G}$, $\bar{G}$, and $2\bar{G}$, where $\bar{G}=3.64$ is the mean zero-shot norm. Across all amplitudes, actual movement direction tracks the commanded heading with a median angular error of $10^\circ$ (mean $16^\circ$; Fig.~\ref{fig3}A), confirming that direction is encoded in the orientation of $\mathbf{G}$ independently of its magnitude. Locomotion speed increases with amplitude (from $1.9\pm0.5$ m/s at $\tfrac{1}{2}\bar{G}$ to $3.5\pm0.7$ m/s at $\bar{G}$) and saturates beyond the training norm due to action clipping by the $\tanh$ nonlinearity ($3.4\pm0.7$ m/s at $2\bar{G}$; Fig.~\ref{fig3}B), suggesting the model operates near the action boundary at the zero-shot norm.

\paragraph{Cold-start adaptation.}
Starting from $\mathbf{G}=\mathbf{0}$, the reward-weighted (RW) update recovers behaviorally useful goal vectors within a handful of episodes: reward per step rises from near zero at episode~1 to $2.7$ r/step by episode~5, reaching $\approx77\%$ of zero-shot performance at episode~10 (Fig.~\ref{fig3}C-D). The $G$-vector trajectories (Fig.~\ref{fig3}E) show adaptation steering $\mathbf{G}$ from the origin toward the zero-shot locus, with directions spreading outward along the ellipse consistent with the commanded headings.

\paragraph{Comparison of adaptation strategies.}
We compare three online update rules: reward-weighted (RW), return-weighted (Q-MC), and the exact TD-MC update derived in the Methods. RW achieves $2.7$ r/step by episode~10, Q-MC $1.9$ r/step, and TD-MC $\approx0$ r/step, failing to adapt. The failure of TD-MC is attributable to two compounding factors: at $\mathbf{G}=\mathbf{0}$ the random policy yields near-zero returns (SNR $=0.18$); and the baseline term $\mathbf{G}^\top\boldsymbol{\psi}_t$ introduces correlated noise once $\mathbf{G}$ departs from zero (gradient SNR $\approx1.2$ at the zero-shot $\mathbf{G}$, with $\sim\!15\%$ of episodes producing gradients of the wrong sign). Removing the baseline and replacing the Monte Carlo return with the instantaneous reward yields the RW rule, which achieves the best cold-start performance despite its simplicity.

We further evaluate the framework on a friction-generalisation task, showing that the G-module spontaneously encodes surface friction along a smooth latent manifold and enables robust transfer to unseen friction conditions; full experimental details and results are presented in Appendix (Fig.~\ref{fig4}).

\section{Discussion}

The results presented here support a coherent picture in which a bilinear gating principle, motivated by the physiology of apical-basal dendritic coincidence detection in L5 pyramidal neurons, provides concrete computational advantages when implemented as a neural network actor-critic. The spiking neuron experiment (Fig.~\ref{fig0}) establishes that burst coincidence is a biologically sufficient substrate for the bilinear product $G_k Y_k$; the subsequent experiments demonstrate that the same principle, instantiated as a standard differentiable actor-critic, yields zero-shot generalisation, interpretable low-dimensional control, and rapid goal adaptation across locomotion tasks. The Unitree Go1 results (Fig.~\ref{fig5}) confirm that these properties generalise beyond a single platform.

\paragraph{Motor synergies and gain modulation.}
The decomposition $\mu = \sum_k G_k Y_k$ connects directly to the muscle synergy framework in motor neuroscience. Electromyographic and force-field studies across species and tasks consistently show that natural movements can be decomposed into a small set of spatially fixed muscle co-activation patterns, with context determining only the time-varying combination coefficients~\cite{mussa1994combining,thoroughman2000learning}. This dimensionality reduction is thought to reflect a fundamental organizational principle of spinal and cortical motor circuits: a reusable basis of motor programs is encoded at the spinal and premotor level, while supraspinal signals modulate which combination is expressed. The policy bases $Y_k(s)$ are the computational analogue of these synergies, encoding state-dependent motor output independently of the current goal. The $G$-module, biologically realized by top-down apical drive to L5 pyramidal neurons, plays the role of the context-dependent weighting signal, consistent with the gain modulation motif in which descending projections multiplicatively scale the output of sensorimotor pathways. The bilinear structure makes this role mechanistically explicit, grounding gain modulation in the coincidence detection properties of L5 dendrites rather than treating it as a phenomenological observation. The decoupled heading-speed structure of the learned $G$-space (Fig.~\ref{fig3}A-B) parallels the known separability of movement direction and vigor in motor cortex populations~\cite{churchland2012neural}, suggesting that the bilinear decomposition and biological motor circuits face a shared computational pressure. The cold-start adaptation results (Fig.~\ref{fig3}C-E) reinforce the synergy analogy: just as animals rapidly recombine existing motor programs in a new context, the reward-weighted update recovers useful behaviour within a handful of episodes by re-estimating $G$ alone, leaving the synergy basis untouched.

\paragraph{Burst fraction as the context signal.}
The burst coincidence experiment reveals a subtlety with direct implications for interpreting neural recordings. Because basal drive is state-dependent rather than goal-dependent, raw spike rate carries little direction information; it is the burst fraction (the proportion of spikes occurring during coincident apical activation) that is the goal-selective signal. This is not a semantic distinction: the same underlying spike is classified as a burst or a tonic event depending solely on whether the apical calcium gate is open. The model predicts that direction selectivity should be substantially stronger in burst events than in total spike counts, a distinction standard firing-rate analyses would miss. This prediction is directly confirmed across two independent datasets (Fig.~\ref{fig0b}): burst fraction DSI is significantly higher than tonic rate DSI in both MC\_Maze ($p<10^{-29}$, $N=177$, monkey Jenkins, Stanford) and MC\_RTT ($p<10^{-16}$, $N=106$, monkey Indy, UCSF), and population activity matrices sorted by burst-fraction preferred direction display a pronounced directional staircase that is absent from the tonic rate matrices in both cases. The consistency across different animals, institutions, and task variants strengthens the interpretation as a general property of M1/PMd coding rather than an artefact of a single recording. The result also suggests that existing electrophysiology studies that report only total spike counts~\cite{churchland2012neural} may systematically underestimate the goal-selectivity of individual neurons.
This interpretation aligns with the view that bursts constitute a distinct, multiplexed communication channel carrying information separate from the mean rate~\cite{naud2018sparse,payeur2021}, with causal evidence that apical dendritic activity gates perception and behaviour~\cite{takahashi2016}, and with learning schemes in which burst-dependent plasticity coordinates credit assignment across cortical layers~\cite{payeur2021,greedy2022}.

\paragraph{Flexible behavioral control via $G$.}
Figures~\ref{fig3}A-B reveal that the $G$-space supports a surprisingly clean factorisation of behavioral degrees of freedom: the \emph{direction} of $\mathbf{G}$ controls the heading of the agent, while its \emph{magnitude} $\|\mathbf{G}\|$ independently controls locomotion speed.
This decoupling is not explicitly trained; the reward signal targets direction only, yet it emerges from the bilinear structure as a natural consequence of learning separable bases.
The result is a practical and intuitive control interface: a user (or a higher-level planner) can specify where the agent should go by rotating $\mathbf{G}$, and how fast by scaling it, without retraining or re-optimizing anything.
The spontaneous decoupling of heading and speed parallels the known separability of movement direction and movement vigor in motor cortex populations~\cite{churchland2012neural}, suggesting that the bilinear structure may reflect a computational pressure that is also at work in biological circuits, rather than being an artefact of the training objective.

The decoupled heading-speed structure is suggestive of partial \emph{monosemanticity} in the $G$-space: behavioral dimensions appear to map onto geometrically separable axes of $\mathbf{G}$ rather than being spread across all components simultaneously.
Full monosemanticity, where each $G_k$ corresponds to exactly one interpretable behavioral feature, is not enforced and not observed here, but the emergent axis-aligned structure suggests that the bilinear decomposition creates internal pressure toward it.
Explicitly encouraging monosemanticity via sparsity or disentanglement regularisers on $\mathbf{G}$ could further sharpen interpretability and is a natural direction for future work.

\paragraph{Scope of zero-shot generalization.}
The zero-shot transfer demonstrated in Fig.~\ref{fig2} relies in part on the regular geometry of the training distribution: all training directions lie on a circle, the linear $g$-layer encodes this manifold as a smooth ellipse in $G$-space, and unseen angles are reached by interpolation within it. The generalization succeeds because intermediate goals remain geometrically close to the training manifold. Whether comparably clean transfer extends to goal families with higher-dimensional or irregular structure, to tasks requiring qualitatively distinct motor programs, or to settings where the goal parameterization is not known in advance, are important open questions. The current results should be interpreted as a demonstration of the structural properties of the bilinear decomposition within a well-defined benchmark family rather than as evidence of general-purpose zero-shot transfer.

\paragraph{Coincidence detection as hardware multiplication and its inductive bias for OOD adaptation.}
A deeper significance of the burst coincidence mechanism lies in its relationship to multiplication. The AND-gate structure of two-compartment neurons implements stochastic multiplication at the hardware level: the burst probability $p_\text{burst} = p_s \times p_d$ is the product of two independent Bernoulli variables, each encoding one input channel. This is not an approximation of multiplication; it is exact in expectation, with stochasticity serving as a Monte Carlo estimator of the product~\cite{vonneumann1956probabilistic,gaines1969stochastic}. As demonstrated in~\cite{capone2025adaptive,capone2023beyond}, multiplicative gates provide the correct inductive bias for estimating multiplicative interactions: architectures with explicit multiplication generalise to out-of-distribution (OOD) inputs more reliably than additive networks that must approximate multiplication through learned nonlinearities, because the architectural prior matches the true functional form. Non-multiplicative networks can in principle learn to approximate multiplication, but require substantially more training examples and exhibit poorer OOD extrapolation. This matters directly for adaptive behaviour: gradient-based plasticity rules, including the reward-weighted update $\Delta G_k \propto r \cdot \psi_k$ used here, are themselves multiplicative operations, and their reliable estimation in new goal contexts benefits from a substrate that implements multiplication natively. The present results extend this argument to goal-directed reinforcement learning: the bilinear decomposition $\mu = \sum_k G_k Y_k$ succeeds precisely because it matches the multiplicative structure of the adaptation problem, yielding zero-shot generalisation and rapid convergence that non-multiplicative baselines do not achieve to the same degree.

\paragraph{Implications for neuromorphic hardware.}
The burst coincidence mechanism also points toward a concrete role for neuromorphic hardware in adaptive control. Multiplications are among the most expensive operations on standard von Neumann architectures, yet they can be implemented at near-zero marginal cost by coincidence circuits in spiking hardware. A population of two-compartment neurons with tunable apical weights constitutes a multiply-accumulate unit whose parameters (the goal vector $G$) can be updated by local Hebbian-like rules without requiring global backpropagation through the circuit. As neuromorphic platforms mature, including spiking-chip designs capable of dense coincidence detection, the bilinear decomposition proposed here offers a concrete blueprint for rapid goal adaptation at the hardware level: reprogramming $G$ amounts to adjusting a low-dimensional apical weight vector, achievable in a handful of synaptic updates rather than a full network re-optimisation. This suggests that the biological solution to flexible motor control may simultaneously be a practical solution for energy-efficient adaptive robotics.

\paragraph{Conceptual comparison with Successor Features and GPI.}
Successor Features (SF)~\cite{barreto2017successor,barreto2018transfer} factorize $Q^\mathbf{w}(\mathbf{s},\mathbf{a}) = \mathbf{w}^\top \boldsymbol{\phi}(\mathbf{s},\mathbf{a})$ and use Generalized Policy Improvement (GPI) to construct a new policy by a componentwise max over a library of task-specific policies.
Our framework shares the linear-in-$G$ value structure but differs in two fundamental ways.
First, $G$ is a continuously adaptable function of goal context rather than a fixed per-task constant, interpolating smoothly between trained objectives as evidenced by the elliptical $G$-locus in Fig.~\ref{fig3}E.
Second, and more importantly, updating $G$ simultaneously updates the policy: actor and critic share $G$, so any change to $G$ immediately reshapes $\mu(\mathbf{s}) = \sum_k G_k Y_k(\mathbf{s})$, making value optimisation and policy improvement a single operation rather than two sequential ones.
This also distinguishes our approach from GPI: rather than a componentwise max over a discrete library, $G$ performs a continuous weighted combination of learned primitives, and the zero-shot generalisation in Fig.~\ref{fig2} is a direct empirical signature of this.
For this reason the appropriate axis of comparison with SF+GPI is not asymptotic reward but \emph{adaptation cost}: after a change in the reward structure, SF+GPI must perform a separate policy-improvement pass (the GPI max, or an actor re-optimisation) to realise the new policy, whereas re-estimating $G$ here updates value and policy in the same operation, at a cost of zero additional policy-improvement steps.
A raw performance benchmark would therefore mis-frame the contribution, which is the elimination of a sequential optimisation stage rather than a gain in reward; the cold-start results (Fig.~\ref{fig3}C-E), where a behaviourally useful policy is recovered within a handful of episodes by adapting $G$ alone, are the direct demonstration of this property.

\paragraph{Value-based adaptation and continual task switching.}
From a mechanistic perspective, adaptation is achieved by updating only $G$ while keeping basis functions fixed. The reward-weighted rule $\Delta G_k \propto r\,\psi_k(s,a)$ propagates reward information directly through the decomposition, yielding rapid behavioral adjustment within a handful of episodes without any policy gradient. In our experiments, TD-based baselines degraded performance, supporting the view that baseline-free return weighting is more reliable in the cold-start regime. A direct consequence is that the framework supports \emph{continual task switching}: when the task changes at deployment time, the agent re-estimates $\mathbf{G}$ online in a few environment steps without touching the pretrained bases. This stands in contrast to fine-tuning, which risks catastrophic forgetting, and to meta-learning, which requires an expensive inner-loop gradient pass. This mirrors the view that rapid behavioral flexibility in animals relies on updating a compact contextual representation while keeping motor programs intact~\cite{wolpert2011principles}, a division of labour that the shared $G$-space makes explicit and mechanistically grounded.

Together, these results establish bilinear gating of motor primitives as a computational principle connecting three levels of analysis: at the neural level, burst coincidence detection in two-compartment neurons implements stochastic multiplication natively, providing the inductive bias for reliable OOD generalisation; at the algorithmic level, the shared bilinear decomposition enables rapid goal adaptation without a separate policy gradient step; and at the engineering level, the same principle offers a blueprint for neuromorphic hardware capable of efficient, reprogrammable motor control. Several directions remain open. The experimental validation is currently confined to simulated locomotion; extending to manipulation tasks, sparse-reward settings, and physical hardware would test the generality of the bilinear advantage, as would a systematic benchmark against the strongest current model-free baselines. On the neuroscience side, the connection to L5 physiology rests on a proof-of-concept simulation: testing the burst-fraction selectivity prediction against motor cortex recordings during goal-switching tasks, and exploring whether $G$-space structure has analogues in prefrontal or premotor representations, would provide direct empirical grounding. A unified model bridging the spiking and rate-coded formulations, potentially enabling neuromorphic deployment, is a natural next step.

\section{Friction Generalisation Experiment}
\label{app:friction}

\begin{figure*}[t!]
\centering
\includegraphics[width=0.9\linewidth]{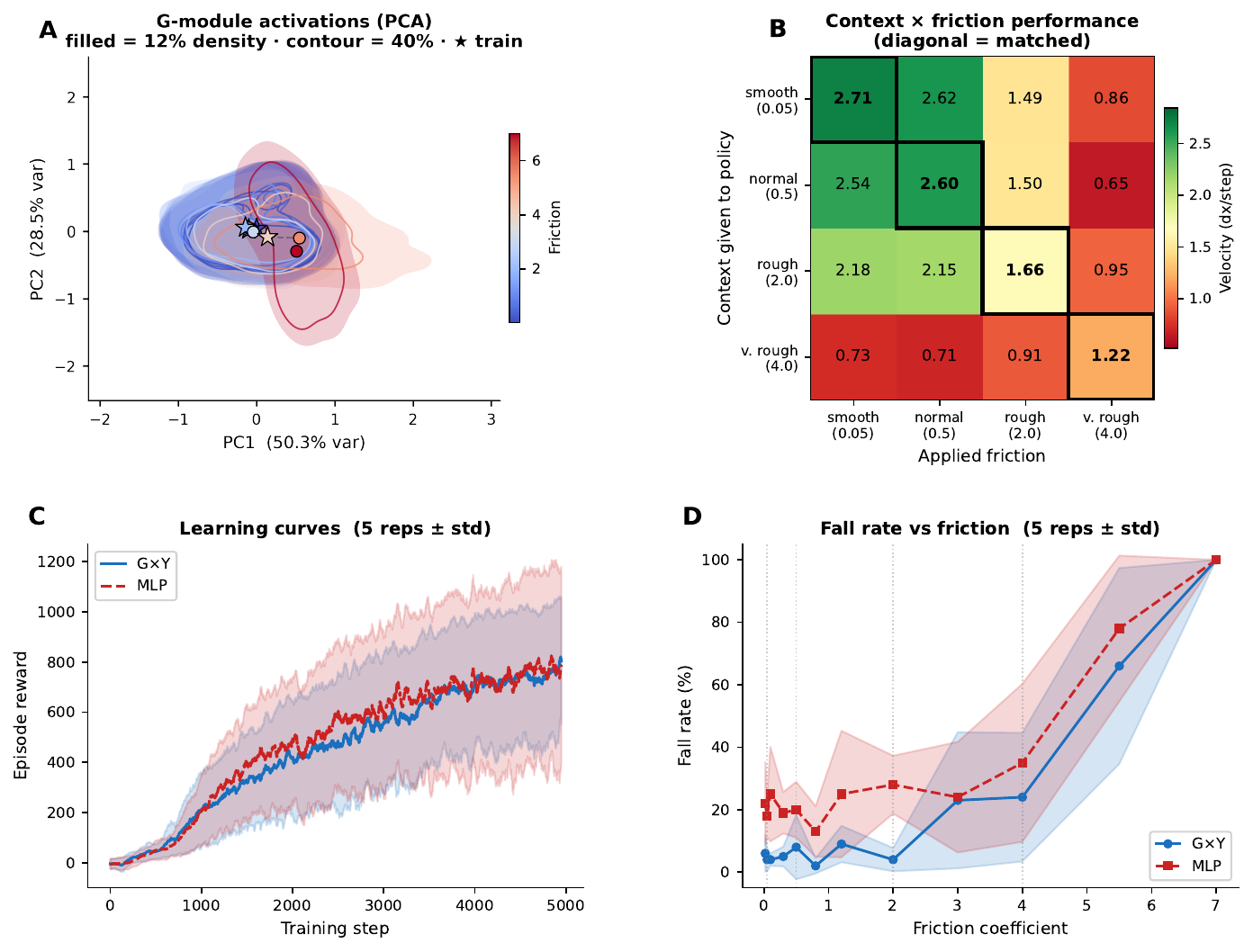}
\caption{\textbf{Multiplicative context-gating enables robust generalisation
        across floor friction conditions.}
        \textbf{(A)} Principal component analysis of the G-module activations
        collected over 100 evaluation episodes per friction level. Each point
        represents the gate vector $\mathbf{g} = G(\mathbf{s}, \mathbf{c})$
        at a single timestep, coloured by friction coefficient
        (blue\,$=$\,low, red\,$=$\,high). Stars denote training frictions;
        circles denote novel test frictions. The smooth trajectory of
        centroids indicates that the gating network encodes friction as a
        continuous latent variable despite receiving only discrete one-hot
        context during training.
        \textbf{(B)} Context-friction performance matrix. Each entry shows
        the mean forward velocity (dx/step) when the agent is given context
        $\mathbf{c}_i$ (row) and deployed on friction $f_j$ (column),
        averaged over 10 episodes. Bold diagonal entries correspond to matched
        context-friction pairs; the diagonal is consistently the highest in
        each row.
        \textbf{(C)} Training curves (mean\,$\pm$\,std over 5 seeds).
        Both G$\times$Y and the MLP baseline achieve comparable final episode
        reward.
        \textbf{(D)} Fall rate as a function of floor friction
        (mean\,$\pm$\,std, 5 seeds, 20 episodes per condition).
        G$\times$Y (blue) maintains near-zero fall rate up to $f = 2.0$ and
        degrades more gracefully than the MLP baseline (red).}
\label{fig4}
\end{figure*}

\paragraph{G-Module activations organise friction along a smooth manifold.}
The PCA of gate activations (Fig.~\ref{fig4}A) reveals that the gating
network spontaneously organises friction conditions along a smooth
one-dimensional manifold in gate space, despite receiving only discrete
one-hot context signals during training.
The centroids of the four training conditions (stars) and the eight novel
test conditions (circles) lie along a continuous trajectory from low-friction
(blue cluster) to high-friction (red cluster), with PC1 alone explaining
$50\%$ of total gate variance.
This indicates that the multiplicative decomposition induces an implicit
metric over friction levels, a property absent by design but emergent
from the training objective.

\paragraph{Matched context is necessary for optimal performance.}
The context-friction matrix (Fig.~\ref{fig4}B) confirms that matched
context is necessary for optimal behaviour.
Diagonal entries, where the provided context matches the actual floor
friction, are consistently higher than off-diagonal entries within the
same column.
The effect is strongest at extreme frictions: providing a smooth context
($\mathbf{e}_1$) on a very rough floor ($f = 4.0$) yields a mean velocity
of $0.04$\,m/s, compared to $1.09$\,m/s with the correct very-rough context:
a 27-fold degradation.
This asymmetry indicates that the agent has learned friction-specific motor
strategies that are genuinely distinct and non-interchangeable.

\paragraph{G$\times$Y learns more stable gaits than the MLP baseline.}
Training curves (Fig.~\ref{fig4}C) show that both G$\times$Y and the
MLP baseline reach comparable final episode reward after 5\,000 iterations,
with overlapping confidence bands across seeds.
The fall rate comparison (Fig.~\ref{fig4}D) exposes a qualitative
difference in the learned behaviour.
At moderate-to-high friction ($f \in [1.0,\,4.0]$), G$\times$Y maintains a
fall rate below $20\%$ across all seeds, whereas the MLP baseline climbs
steeply, reaching $60$-$100\%$ at $f = 5.5$.
This difference does not manifest as a large velocity gap (the MLP moves
fast when it does not fall) but accumulates as shorter episodes and lower
total reward.
We interpret this as evidence that the multiplicative gating mechanism
promotes conservative, stable gaits: the G-module modulates action magnitude
in a friction-dependent manner, effectively reducing destabilising motor
commands on rough terrain.

\paragraph{Environment.}
We used the MuJoCo Ant-v4 locomotion task~\cite{todorov2012mujoco} as
implemented in OpenAI Gym~\cite{brockman2016openai}.
The ant must walk forward on a flat floor; an episode terminates if the
torso height falls below a fixed threshold.
We modified floor friction at runtime by setting
\texttt{model.geom\_friction[0,\,0]} to one of four values
$f \in \{0.05,\,0.5,\,2.0,\,4.0\}$, corresponding to smooth, normal, rough,
and very rough surfaces respectively.

\paragraph{Context encoding.}
During training, each episode is assigned a friction level sampled uniformly
from the four training conditions.
The agent receives a four-dimensional one-hot vector
$\mathbf{c} \in \{\mathbf{e}_1, \mathbf{e}_2, \mathbf{e}_3, \mathbf{e}_4\}$
indicating the current friction, appended to the 27-dimensional
proprioceptive observation, yielding a 31-dimensional input.
The friction level is switched every 100 timesteps within each episode.

During evaluation on novel friction values, the context vector is computed
as a linear interpolation between the two nearest training one-hot vectors:
\begin{equation}
    \mathbf{c}(f) = (1 - w)\,\mathbf{e}_i + w\,\mathbf{e}_{i+1},
    \qquad
    w = \frac{f - f_i}{f_{i+1} - f_i},
    \quad f \in [f_i,\, f_{i+1}].
\end{equation}
For friction values outside the training range, the nearest endpoint
one-hot vector is used.

\paragraph{Architecture.}
The G$\times$Y actor decomposes the policy as
\begin{equation}
    \boldsymbol{\mu}(\mathbf{o}) =
        \sum_{k=1}^{K} G_k(\mathbf{o})\; \mathbf{Y}_k(\mathbf{o}),
    \qquad
    \mathbf{a} = \tanh\!\left(
        \boldsymbol{\mu}(\mathbf{o}) +
        \boldsymbol{\sigma}(\mathbf{o})\,\boldsymbol{\epsilon}
    \right),
\end{equation}
where $\mathbf{o} = [\mathbf{s};\,\mathbf{c}]$ is the augmented observation,
$G: \mathbb{R}^{31} \to \mathbb{R}^K$ is a single linear layer producing
scalar gates ($K\!=\!4$), $\{\mathbf{Y}_k: \mathbb{R}^{31} \to \mathbb{R}^8\}$
are $K$ linear action modules, and
$\boldsymbol{\sigma}(\mathbf{o}) = \exp(\text{clip}(W_\sigma \mathbf{o},\,
\sigma_{\min},\, \sigma_{\max}))$ is produced by a separate linear head.
The baseline actor is a two-hidden-layer MLP (256 units, ReLU) with identical
input and output dimensionality.
Both actors share the same twin-critic architecture (256-unit MLP).

\paragraph{Training.}
All models were trained with Soft Actor-Critic~\cite{haarnoja2018soft}
with entropy coefficient $\alpha$ auto-tuned, discount $\gamma = 0.99$,
target smoothing $\tau = 0.005$, batch size 256, and learning rate
$3 \times 10^{-4}$ for all networks.
Each model was trained for 5\,000 environment iterations of 400 steps each,
with 50 gradient updates per iteration after a replay buffer warm-up of
256 transitions.
We ran 5 independent seeds per model.

\paragraph{Evaluation.}
Generalisation was assessed on 12 friction values spanning and extrapolating
beyond the training range:
$f \in \{0.02,\,0.05,\,0.1,\,0.3,\,0.5,\,0.8,\,1.2,\,2.0,\,3.0,\,4.0,\,
5.5,\,7.0\}$.
For each friction we ran 20 episodes (400 steps maximum) per seed and
report mean forward velocity (dx/step, averaged over all timesteps and
episodes), fall rate (fraction of episodes terminated by the health
condition), and mean episode length.
For the context-friction matrix (Fig.~\ref{fig4}B) we used the
representative seed and ran 10 episodes per cell.

\section{Gating Representations in SAC Locomotion}
\label{app:gating}

\begin{figure*}[t!]
\centering
\includegraphics[width=\linewidth]{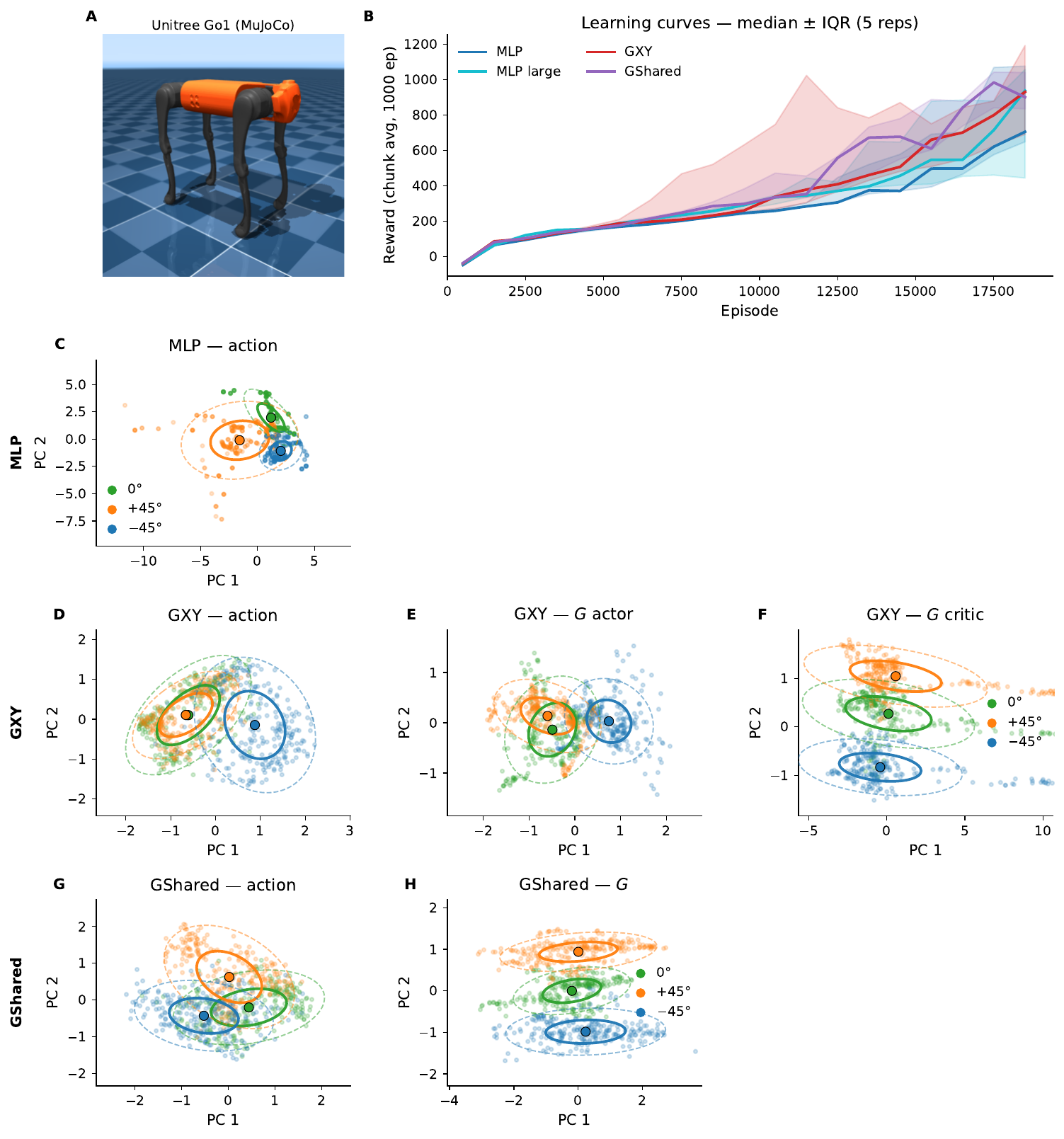}
\caption{\textbf{Gating representations in SAC locomotion on the Unitree Go1 quadruped.}
        \textbf{(a)} MuJoCo simulation environment.
        The robot is rewarded for walking in a direction
        $\theta \in \{-45^{\circ},\,0^{\circ},\,+45^{\circ}\}$
        (encoded as $[\cos\theta,\sin\theta]$ appended to the observation),
        which changes randomly every 500 steps within each 1500-step episode.
        \textbf{(b)} Learning curves for four architectures
        (median\,$\pm$\,IQR over 5 independent runs, averaged in chunks of
        1000 episodes): MLP baseline, MLP-large (parameter-matched critic),
        GXY (independent gating in actor and critic), and GShares (shared $G$
        layer between actor and critic).
        \textbf{(c-h)} PC1 vs.\ PC2 projections of neural representations
        collected across 15 rollouts per direction (colours denote target
        direction). Gaussian ellipses mark the $1\sigma$ (solid) and $2\sigma$
        (dashed) contours of each direction's distribution.
        \textbf{(c)} MLP action space shows overlapping distributions across
        directions.
        \textbf{(d-f)} GXY actor action space, actor gating variable $G$,
        and critic gating variable $G$: the $G$ spaces of both actor and
        critic exhibit clear directional clustering, indicating that the gating
        mechanism spontaneously organises into direction-selective modules.
        \textbf{(g-h)} GShares action space and shared $G$: the shared
        gating layer similarly develops direction-selective activity despite
        being trained only through the critic loss.}
\label{fig5}
\end{figure*}

This note reports the full methods and results for the Unitree Go1 directional
locomotion experiment.

\paragraph{Environment.}
We use the Unitree Go1 quadruped simulated in MuJoCo.
The observation is the concatenation of joint positions and velocities
($\mathbf{q} \in \mathbb{R}^{37}$) and a 2D unit vector encoding the target
direction $\theta$:
$\mathbf{o} = [\mathbf{q};\, \cos\theta,\, \sin\theta] \in \mathbb{R}^{39}$.
The action space consists of 12 joint torques clipped to $[-0.863, 0.863]$\,N{\textperiodcentered}m.
The instantaneous reward is the projected velocity along the target direction:
$r_t = \dot{x}_t \cos\theta + \dot{y}_t \sin\theta$.
An episode terminates early if pitch or roll exceeds $15^{\circ}$, yaw exceeds $90^{\circ}$,
or the base height falls below $0.2$\,m.

\paragraph{Task.}
Each episode lasts 1500 simulation steps.
The target direction is drawn uniformly from $\{-45^{\circ},\, 0^{\circ},\, +45^{\circ}\}$ and
switches every 500 steps, requiring the agent to adapt its gait within an
episode without a reset.

\paragraph{Algorithm.}
All agents are trained with SAC with entropy regularisation
(target entropy $= -|\mathcal{A}|$, $\gamma = 0.99$, $\tau = 0.005$,
learning rate $3 \times 10^{-4}$, batch size 256, 20 gradient updates per
episode).
We run 5 independent repetitions of 50\,000 episodes each.

\paragraph{Architectures.}
We compare four agents:
\begin{itemize}
  \item \textbf{MLP}: standard two-hidden-layer actor (256 units each) and
        two-hidden-layer critic taking the full observation and action as input.
  \item \textbf{MLP-large}: identical actor; critic extended to three hidden
        layers, matching the parameter count of GXY (${\approx}372$k parameters).
  \item \textbf{GXY}: mixture-of-experts actor and critic with $k = 4$ gating
        units. The actor computes
        $\hat{\mathbf{a}} = \sum_{i=1}^{k} g_i(\mathbf{o})\, Y_i(\phi(\mathbf{o}))$,
        where $\phi$ is a shared two-layer trunk, $g_i = [\mathbf{g}(\mathbf{o})]_i$
        are scalar gates produced by a linear layer applied directly to the
        observation, and $Y_i$ are per-expert linear heads. The direction vector is
        excluded from the critic's trunk input (but retained in $\mathbf{g}$).
  \item \textbf{GShares}: identical to GXY, but the actor's gate layer is
        replaced by a reference to the critic's gate layer
        ($\mathbf{g}_{\text{actor}} \equiv \mathbf{g}_{\text{critic}_1}$);
        the actor optimiser excludes the shared gate parameters, so gates are
        trained exclusively by the critic loss.
\end{itemize}

\paragraph{Analysis.}
After training, we select the best repetition per architecture (highest mean
reward over the last 500 episodes) and run 15 evaluation rollouts per direction
(3000 steps, no early termination for data collection).
We collect per-timestep action vectors and gate activations
$\mathbf{g} \in \mathbb{R}^{4}$, pool all timesteps across trials, fit PCA on
the pooled data, and project each direction's data onto the first two principal
components.

\paragraph{Learning.}
All four architectures learn to walk in the commanded direction
(Fig.~\ref{fig5}b).
MLP-large does not improve over the MLP baseline despite the larger critic,
confirming that raw capacity is not the bottleneck.
GXY and GShares reach comparable asymptotic performance, with GShares showing
slightly higher variance across seeds, consistent with its more constrained
gating (driven only by the critic signal).

\paragraph{Direction-selective gating.}
The PC analysis (Fig.~\ref{fig5}c-h) reveals a qualitative difference between
architectures.
In the MLP, action distributions for the three directions overlap substantially
in PC1-PC2 space (Fig.~\ref{fig5}c), indicating that the network does not
develop clearly separable internal representations per direction.
By contrast, both GXY and GShares develop strongly direction-selective gate
activations: the $\mathbf{g}$ distributions for different target directions
occupy well-separated clusters in PC space
(Fig.~\ref{fig5}e-f,\,h), while their action distributions remain moderately
overlapping (Fig.~\ref{fig5}d,\,g).
This dissociation suggests that the gating layer absorbs directional context,
freeing the shared trunk to compute direction-invariant motor primitives that
are then linearly combined by the gates.

\paragraph{Shared vs.\ independent gating.}
In GShares, the gate is trained only by the critic loss, yet it achieves a
degree of direction selectivity comparable to GXY
(Fig.~\ref{fig5}h vs.\ Fig.~\ref{fig5}e).
This indicates that value prediction alone is sufficient to drive the emergence
of directional structure in the gating layer, without requiring an explicit
actor gradient through the gate.
The result is consistent with the interpretation that the critic, which must
predict direction-dependent returns, learns a gate representation that
partitions state space by commanded direction, a representation the actor
then exploits through its expert heads.

\section{Methods}

\subsection{Burst Neuron Model (V-prob Variant)}

Each output unit of the actor network is modelled as a two-compartment integrate-and-fire neuron. The somatic voltage $V_s$ and dendritic voltage $V_d$ evolve as leaky integrators with shared time constant $\tau_s = \tau_d = 5$\,ms:
\begin{equation}
\dot{V}_s = \frac{1}{\tau_s}\!\left(-V_s + I_\text{bas}\right), \qquad
\dot{V}_d = \frac{1}{\tau_d}\!\left(-V_d + I_\text{api}\right).
\end{equation}
The somatic current is $I_\text{bas} = I_\text{offset} + I_\text{scale}\cdot\tanh(y)$, with $I_\text{offset}=0.5$ and $I_\text{scale}=1.3$, where $y$ is the network readout. The apical current is $I_\text{api} = g \cdot I_\text{API\_SCALE}$ with $I_\text{API\_SCALE}=4.0$, where $g\in\{0,1\}$ is the one-hot goal signal. The increased apical drive compared to earlier formulations ensures the dendritic gate is driven well above saturation ($p_d \approx 1$) whenever the corresponding goal is active.

\paragraph{Voltage-based burst probability.}
At the start of each 15-ms action window, a burst event is stochastically initiated with probability $p_s(V_s) \times p_d(V_d)$, where
\begin{equation}
p_s = \sigma\!\left(\frac{V_s - \theta_s}{\beta_s}\right), \quad
p_d = \sigma\!\left(\frac{V_d - \theta_d}{\beta_d}\right),
\end{equation}
with $\theta_s=1.0$, $\beta_s=0.5$, $\theta_d=0.75$, $\beta_d=0.2$. If a burst is initiated, $V_d$ is reset to zero and the neuron enters a fast-spiking regime (refractory $T_\text{ref,burst}=1.5$\,ms, reset $V_s=0.9$) for 15\,ms, followed by a dendritic refractory period of $T_\text{ref,d}=30$\,ms. The action signal is the time-averaged product over the $N_\text{steps}=15$ timesteps of each window:
\begin{equation}
p_\text{burst} = \frac{1}{N_\text{steps}} \sum_{t=1}^{N_\text{steps}} p_s\!\left(V_s(t)\right)\cdot p_d\!\left(V_d(t)\right).
\end{equation}
This voltage-based estimate faithfully reflects the instantaneous membrane state throughout the window rather than a fixed equilibrium approximation.

\paragraph{Action encoding.}
The motor command is
\begin{equation}
\mu = P_\text{scale} \cdot p_\text{burst} - 1, \quad P_\text{scale} = 7,
\end{equation}
with $\mu$ clipped to $[-1,1]$ before submission to the environment. The neutral operating point is $p_\text{burst}^\star = 1/P_\text{scale} \approx 0.143$, giving $\mu=0$. The policy gradient through the burst probability is
\begin{equation}
\frac{\partial \mu}{\partial y} = P_\text{scale} \cdot p_\text{burst}(1-p_\text{burst}) \cdot I_\text{scale} \cdot \operatorname{sech}^2(y)\,/\,\beta_s,
\end{equation}
and synaptic weights are updated via a filtered eligibility trace with a homeostatic baseline.

\paragraph{Training (burst experiment).}
The agent was trained for 3500 episodes on a four-goal variant of MuJoCo Ant-v4. Within each episode (400 steps, each corresponding to 15\,ms of burst-neuron simulation), the active goal context cycled through the four cardinal directions ($+X$, $-X$, $+Y$, $-Y$) every 100 steps (1500\,ms), following a random permutation. Rewards were the instantaneous velocity in the target direction. Output weights were updated by the Adam optimiser with $\alpha=8\times10^{-5}$.

\subsection{Pretraining via Hierarchical Soft Actor-Critic}

We pretrain the policy using a goal-conditioned Soft Actor-Critic (SAC~\cite{haarnoja2018soft}) with a two-level hierarchical structure. The complete model defined below, comprising the bilinear actor, the factored critic, and their joint optimisation, constitutes a soft actor-critic trained with twin soft Q-functions, entropy-regularised actor updates, and target networks, as specified under \emph{Joint training}.

\paragraph{High-level context mapping.}
A task context $\mathbf{c} = (\cos\theta, \sin\theta)$ encoding the desired locomotion direction is mapped to a latent goal vector through a linear, bias-free projection:
\begin{equation}
\mathbf{G} = W_g\,\mathbf{c} \;\in\; \mathbb{R}^K.
\label{eq:g_map}
\end{equation}
The omission of a bias term is a deliberate design choice: it constrains the image of the context manifold to pass through the origin in $G$-space, ensuring that $\mathbf{G}=\mathbf{0}$ is a valid neutral state from which cold-start adaptation can be launched without prior knowledge of the task.

\paragraph{Low-level policy via motor primitives.}
$\mathbf{G}$ weights a set of $K$ learned motor primitives $\{Y_j\}_{j=1}^K$, each a nonlinear function of the proprioceptive state $\mathbf{s}$, to produce the action mean:
\begin{equation}
\pi(\mathbf{s}, \mathbf{G}) = \tanh\!\left(\sum_{j=1}^K G_j\, Y_j(\mathbf{s})\right).
\label{eq:policy_bilinear}
\end{equation}
This induces a \emph{bilinear} structure (linear in $Y_j$ for fixed $\mathbf{G}$, and linear in $\mathbf{G}$ for fixed state features) that disentangles \emph{what actions are available} (encoded by $Y_j$) from \emph{how strongly they are expressed} (controlled by $G_j$).

\paragraph{Factored critic.}
The critic adopts the same factored structure, expressing the action-value function as a linear combination of learned basis functions $\boldsymbol{\psi}(\mathbf{s},\mathbf{a})\in\mathbb{R}^K$:
\begin{equation}
Q(\mathbf{s}, \mathbf{a}, \mathbf{G}) = \mathbf{G}^\top \boldsymbol{\psi}(\mathbf{s}, \mathbf{a}).
\end{equation}
This mirrors the successor feature framework~\cite{barreto2017successor}, with $\boldsymbol{\psi}$ playing the role of task-agnostic value basis functions and $\mathbf{G}$ the role of task-specific reward weights. Because actor and critic share the same $\mathbf{G}$, any update to $\mathbf{G}$ simultaneously adjusts both value estimates and the executed policy.

\paragraph{Joint training.}
The full model, comprising $W_g$, the primitive networks $\{Y_j\}$, the basis functions $\boldsymbol{\psi}$, and a log-standard deviation head, is trained jointly using Soft Actor-Critic (SAC~\cite{haarnoja2018soft}) across $N=8$ equally spaced directions $\theta\in\{0^\circ, 45^\circ, \ldots, 315^\circ\}$, with direction sampled uniformly at the start of each episode. Two critics $Q_1, Q_2$ are maintained with target networks updated via exponential moving average. Critic updates minimise the squared Bellman error:
\begin{equation}
\mathcal{L}_{Q_i} = \mathbb{E}_{(s,a,r,s')} \Big[ \big( Q_i(\mathbf{s},\mathbf{a},\mathbf{G}) - y \big)^2 \Big], \quad y = r + \gamma \min_j Q_j(\mathbf{s}',\mathbf{a}',\mathbf{G}'),
\end{equation}
and actor updates follow the soft policy gradient:
\begin{equation}
\mathcal{L}_\pi = \mathbb{E}_\mathbf{s} \big[ \alpha \log \pi(\mathbf{a}|\mathbf{s}) - \min_j Q_j(\mathbf{s},\mathbf{a},\mathbf{G}) \big].
\end{equation}
After pretraining, $W_g$ and the primitive networks $\{Y_j\}$ are frozen. At test time, $\mathbf{G}$ is either computed zero-shot via Eq.~\eqref{eq:g_map} or adapted online from $\mathbf{G}=\mathbf{0}$ using only the interaction signal, as described below.

\subsection{Zero-Shot Adaptation}

Before any online update, we evaluate zero-shot transfer by freezing all learned parameters and directly conditioning the policy on a new goal descriptor $g^*$. In this setting, the gating network produces coefficients $G(s,g^*)$ with a single forward pass, and the policy is executed without gradient steps or replay-buffer updates:
\[
\pi_{\text{ZS}}(a|s,g^*) = \sum_{k=1}^{K} G_k(s,g^*)\,Y_k(a|s).
\]
This protocol measures how well the pretrained bases and shared coefficient structure generalize to unseen directions/tasks purely through interpolation in the learned latent space. We report zero-shot performance using return and trajectory alignment metrics before enabling any adaptation dynamics.

\subsection{Online Adaptation of the Goal Vector}

The factorised critic $Q(s,a,\mathbf{G}) = \mathbf{G}^\top \boldsymbol{\psi}(s,a)$ provides a natural mechanism for fast test-time adaptation: given a new task, we need only update the low-dimensional vector $\mathbf{G}\in\mathbb{R}^K$ while keeping the basis functions $\boldsymbol{\psi}$ fixed.

\paragraph{Exact update.}
Given trajectories collected under the current policy, the optimal $\mathbf{G}$ minimises the mean-squared error between the value estimate and the realised Monte Carlo returns:
\[
  \mathcal{L}(\mathbf{G}) = \tfrac{1}{2}\,\mathbb{E}_t\!\left[\bigl(R_t - \mathbf{G}^\top\boldsymbol{\psi}_t\bigr)^2\right],
\]
where $R_t = \sum_{s \geq t} \gamma^{s-t} r_s$ is the discounted return from step $t$ and $\boldsymbol{\psi}_t = \boldsymbol{\psi}(s_t,a_t)$. The gradient update is
\[
  \Delta\mathbf{G} = \eta\,\frac{1}{T}\sum_t \bigl(R_t - \mathbf{G}^\top\boldsymbol{\psi}_t\bigr)\,\boldsymbol{\psi}_t.
\]
This minimises the projection error of the true return onto the learned basis, fully exploiting the linear structure of the critic.

\paragraph{Removing the value baseline.}
The term $\mathbf{G}^\top\boldsymbol{\psi}_t$ acts as a state-dependent baseline that in principle reduces variance. In practice, however, $\boldsymbol{\psi}_t$ is estimated by a critic trained jointly with the actor and does not in general satisfy $Q^\pi = \mathbf{G}^\top\boldsymbol{\psi}$ exactly for arbitrary $\mathbf{G}$. The resulting mismatch introduces correlated noise, and the baseline can point in a direction misaligned with the behaviourally optimal $\mathbf{G}$. Concretely, the signal-to-noise ratio of the TD-MC gradient near the zero-shot $\mathbf{G}$ is only $\approx 1.2$, with a substantial fraction of episodes producing gradients of the wrong sign. Removing the baseline eliminates this instability and yields the simpler return-weighted update:
\[
  \Delta\mathbf{G} = \eta\,\frac{1}{T}\sum_t R_t\,\boldsymbol{\psi}_t.
\]

\paragraph{Online approximation.}
Computing $R_t$ requires a full episode rollout followed by a backward pass, precluding within-episode adaptation. For online use we replace the Monte Carlo return with the instantaneous reward $r_t$, giving the reward-weighted (RW) update applied after every step:
\[
  \Delta\mathbf{G}_t = \eta\, r_t\,\boldsymbol{\psi}_t.
\]
This update is computationally trivial, requires no reward buffering, and allows $\mathbf{G}$ to be refined continuously as the agent interacts. After each episode we project $\mathbf{G}$ onto the ball $\|\mathbf{G}\| \leq G_{\max}$ to prevent unbounded growth.

\subsection{Experimental Setup}

We evaluate our approach on the MuJoCo \texttt{Ant-v4} environment~\cite{todorov2012mujoco} as implemented in OpenAI Gym~\cite{brockman2016openai}. The agent is trained to move in a given target direction, encoded as a 2D vector appended to the observation. During training, directions are cycled every 100 steps through eight angles (four cardinal and four diagonal), ensuring exposure to multiple behavioral modes. Episodes last 800 steps, and reward is defined as forward progress along the target direction minus a small penalty for orthogonal movement:
\begin{equation}
r = \Delta x \cos\theta + \Delta y \sin\theta - 0.1\,|\Delta x \sin\theta - \Delta y \cos\theta|.
\end{equation}

We employ a replay buffer of size $10^5$ and train both actor and critic using Adam with learning rate $3\times10^{-4}$. Each training iteration samples a batch of transitions and performs standard SAC updates, while online G-space adaptation is applied during rollout to demonstrate rapid directional modulation.

\subsection{Multi-Session Burst-Fraction Analysis}

\paragraph{Datasets.}
We analysed 12 recording sessions from three macaque monkeys across two institutions (Table~\ref{tab:sessions}).
Five sessions from monkey Jenkins and six from monkey Nitschke were obtained from DANDI~000070~\cite{churchland2012neural}, the Shenoy lab's full data release associated with multiple publications on motor cortex population dynamics.
One session from monkey Indy was obtained from the MC\_RTT benchmark (DANDI~000129~\cite{pei2021nlb}), recorded at UCSF during a self-paced random target reaching task.
The two benchmark sessions (MC\_Maze DANDI~000128 and MC\_RTT DANDI~000129) are analysed in Fig.~\ref{fig0b}; the remaining 10 sessions constitute the replication set.

\paragraph{Spike extraction and burst detection.}
For each trial we extracted spikes in the 500\,ms window following the go cue (DANDI~000070) or trial onset (DANDI~000129).
A spike was classified as a burst spike if the interval to its immediately preceding spike was $<$10\,ms (ISI threshold); all remaining spikes were classified as tonic.
Burst fraction per neuron per direction was computed as the ratio of burst spikes to total spikes, averaged across trials.
Tonic rate was computed as tonic spike count divided by window duration (0.5\,s), averaged across trials.

\paragraph{Direction binning.}
For DANDI~000070, reach direction was derived from the \texttt{hit\_target\_position} field of each trial, binned to the nearest 45$^\circ$ sector.
For DANDI~000129, direction was computed as the angle of the cursor-to-target vector at trial onset, using only trials where the cursor-to-target distance exceeded 20\,mm to exclude trials where the monkey had already reached the target.

\paragraph{Quality filters.}
A neuron was included if it had $\geq$\texttt{MIN\_TRIALS} trials in every analysed direction bin and a mean combined response (tonic rate + burst fraction) $> 0.2$.
For monkey Jenkins and monkey Indy, \texttt{MIN\_TRIALS}$=5$; for monkey Nitschke, some sessions had severely unbalanced direction sampling due to maze barrier configurations (as few as 5 trials in one of 8 bins), which biased DSI estimates.
For Nitschke, we therefore applied \texttt{MIN\_TRIALS}$=30$ and excluded direction bins with fewer than 30 trials before running the neuron filter, retaining between 4 and 7 of the 8 direction bins depending on the session.
One Nitschke session (20100923) returned N$=0$ neurons under \texttt{MIN\_TRIALS}$=5$ because the least-sampled direction had fewer than 5 trials for all neurons; it qualified under the \texttt{MIN\_TRIALS}$=30$ bin-filtering scheme (7/8 bins retained, $N=192$).

\paragraph{Statistical test and controls.}
For each session we computed the Wilcoxon signed-rank test (one-sided, burst fraction DSI $>$ tonic rate DSI) and the fraction of neurons above the identity diagonal in the DSI scatter.
To verify that the population-level difference was not driven by a small number of outlier neurons, we computed a bootstrap 95\% confidence interval on the percentage above diagonal by resampling neurons with replacement (200 iterations).
To verify that higher burst-fraction DSI is not simply inherited from directional rate modulation, we also computed DSI on the \emph{rate-residual} burst fraction: for each neuron, the burst fraction after regressing out the linear effect of tonic rate across directions.

\clearpage
\bibliographystyle{naturemag}
\bibliography{references}

\end{document}